\documentstyle[11pt,newpasp,twoside]{article} 
\def\edcomment#1{\iffalse\marginpar{\raggedright\sl#1\/}\else\relax\fi} 
\reversemarginpar 

\begin{document} 
\title{When you wish upon a star: Future developments 
in astronomical VLBI.}

 \author{M.A. Garrett} 
\affil{Joint Institute for VLBI in Europe, Postbus 2, 7990~AA
  Dwingeloo, The Netherlands} 

\begin{abstract} 
  In this paper, I present the likely technological development of
  VLBI, and its impact on the astronomical community over the next 1-5
  years.  VLBI is currently poised to take advantage of the rapid
  development in commercial off-the-shelf (COTS) PC-based products. The
  imminent deployment of disk-based recording systems will enable Gbps
  data rates to be achieved routinely by both cm and mm-VLBI networks.
  This, together with anticipated improvements in collecting area,
  receiver systems and coherence time is set to transform the
  performance of VLBI in terms of both baseline and image noise
  sensitivity. At the same time the feasibility of using fibre based
  communication networks as the basis for production, real-time VLBI
  networks will begin. Fantastic new correlator output data rates, and
  the ability to deal with these via powerful PC clusters promises to
  expand the typical VLBI field-of-view to scales previously reserved
  for connected, short baseline interferometers. By simultaneously
  sampling the {\it summed} response of all compact radio sources
  within (and indeed beyond) the half-power point of the VLBI telescope
  primary beam, simple self-calibration of the target field will {\it
    always} be possible at frequencies below a few GHz. Unbiased,
  broad-band continuum surveys will be conducted over huge areas of
  sky, and (redshifted) spectral-features will be detected too. By the
  end of the decade the microJy radio sky will be accessible to VLBI:
  dozens of sources will be simultaneosuly observed, correlated,
  detected and fully analysed all within the same day.
\end{abstract} 

\section{Introduction} 

In this paper I review the possible future development of astronomical
Very Long Baseline Interferometry (VLBI). My analysis is very much
user-driven --- the terms of the review being made via a user-based
``wish list''. The focus falls on several areas that are important to
the astronomical community, and which are likely to be addressed by the
new technologies presented throughout this volume. A similar paper is
presented by Schl\"{u}ter (this volume) regarding likely advances in
Geodetic VLBI activities. 

Naturally, some of the points addressed here represent areas in which I
am personally interested or involved. While these may receive extra
attention, I have tried to maintain a reasonable balance throughout the
text. The astronomical wish-list thus
includes:
\begin{itemize}
\item vastly improved raw baseline and image sensitivity, 
\item extended frequency coverage \& agility, 
\item enhanced image fidelity, 
\item more flexible, robust and reliable VLBI data, including auxiliary
  (calibration) data,
\item sharper spatial and spectral resolution,   
\item fantastic new correlator capacities \& sufficiently capable
offline computing resources (including GRIDs) to analyse the data, 
\item access to a significantly {\it deeper} \& {\it wider} field-of-view,
\item the introduction of automatic data calibration and new analysis
  techniques.
\end{itemize} 

I will review each of these areas in turn, and consider the prospects
for possible improvements. It should be noted that this review is made
at a crucial juncture in the development of VLBI -- in particular the
rapid development of commercial off-the-shelf (COTS) PC-based products
are expected to make a substantial impact in many areas of VLBI that
have previously enjoyed only incremental advances.  These new
developments include: flexible digital signal processing, disk-based
data recording, data transfer via transnational, broad-band, internet
communication networks and capable (offline) computing resources via
(Linux) PC clusters. Over the next decade, the opportunities for making
substantial progress are excellent.

\begin{figure}
\vspace{8.9cm}  
\includegraphics{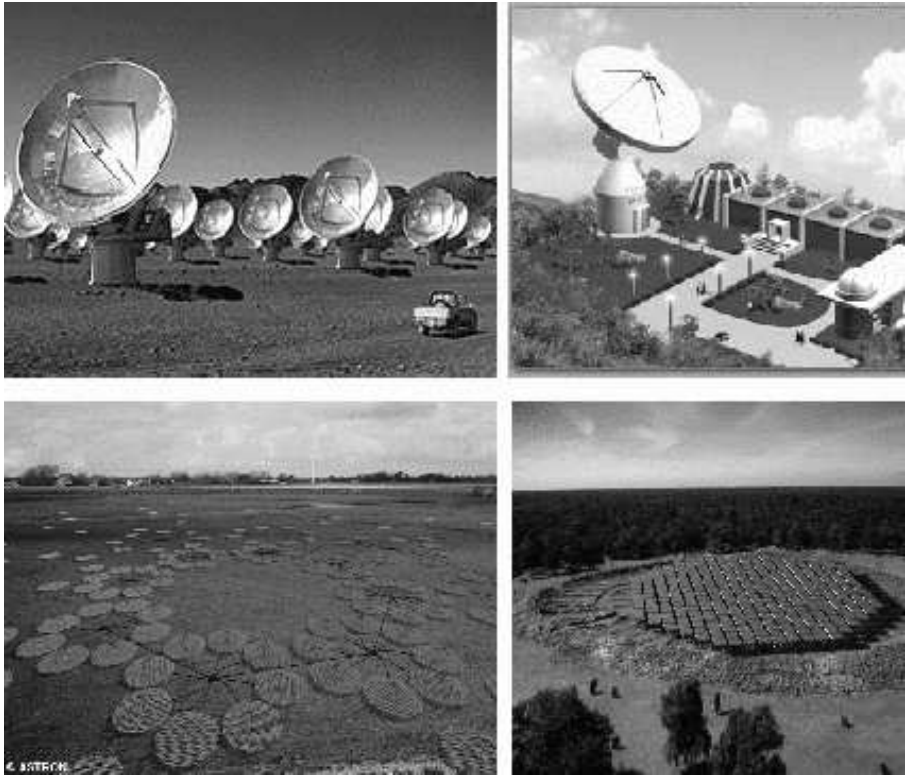}
\caption{Large gains in collecting area can be expected in mm-VLBI with
  the incorporation of ALMA (top left) and other new mm-capable
  telescopes (e.g. the KVN top right). The construction of LOFAR
  (bottom left) is expected to lead to a resurgence in interest in
  meter-VLBI. On longer times-scales, the SKA (bottom right) will lead
  (either directly or indirectly) to a substantial
  increase in the collecting area available  to cm-VLBI. }
\label{area} 
\end{figure}

\section{VLBI Sensitivity} 

In VLBI (and radio interferometry in general) the image and baseline
sensitivity ($\sigma_{I}$ and $\sigma_{B}$ respectively) are dependent
on several different parameters. Two of these are specific to the
antenna - the effective collecting area ($A_{e}$, m$^{2}$) and the
antenna system noise ($T_{sys}$, Kelvin). In addition, sensitivity to
broadband continuum radio emission is determined by the output
fluctuations of the receiver, and so the signal-to-noise ratio is
proportional to the square root of both the spanned observing bandwidth
($\Delta \nu$ Hz) and the total (coherent) integration time
($\tau$, seconds). In SI units the $1\sigma$ baseline sensitivity is
given by: 

\begin{equation}
\sigma_{B} =
\frac{2k\eta_{b}}{\sqrt{2 \Delta \nu \tau}}\sqrt{\frac{T_{sys1}T_{sys2}}{A_{e1}A_{e2}}}
\end{equation}    

\noindent
where $k$ is Boltzmanns's constant \& $\eta_{b}$ accounts for various
losses (see Walker et al. 1988 for more details). Note that the
baseline sensitivity improves (i.e. $\sigma_{B}$ decreases)
proportionally with the geometric average of the effective area of the
two telescopes, and inversely proportional to the geometrical average
of their system temperatures.

Similarly, the $1\sigma$ r.m.s. image noise (assuming optimal data
weighting -- natural weighting) is
given by:  
\begin{equation} 
\sigma_{I} = \frac{2k\eta_{b}}{\sqrt{\Delta \nu \tau}}
\sqrt{\frac{1}{(\sum_{i}\frac{A_{i}}{T_{sysi}})^{2} - \sum_{i}(\frac{A_{i}}{T_{sysi}})^{2}}} 
\end{equation} 

It is clear from these expressions that improvements in sensitivity can
be obtained via four (largely) independent parameters: (i) increased
collecting area, (ii) lower noise receiver systems, (iii) longer
coherence times \& (iv) larger total observing bandwidths.

\subsection{Collecting Area} 

Any increase in collecting area is expected to be incremental at cm
wavelengths but over the next decade, significant gains could be made
at both higher and lower frequencies (see Figure 1). In particular,
mm-VLBI has much to gain by incorporating ALMA within existing networks
(see Alef et al.  this volume) and similarly low-frequency (meter
wavelength) VLBI is bound to be stimulated by the construction of the
Low Frequency Array (LOFAR). mm-VLBI (and to a lesser extent cm-VLBI)
will also benefit from the addition of new telescopes in Asia (VERA and
the KVN -- see Minh et al., Kobayashi et al. and Sassao et al., this
volume) and in Europe (the Yebes-OAN 40-m telescope and the IRA 64-m
Sardinian Radio Telescope). A significant increase in collecting area
for cm-VLBI will have to await the construction of the next
generation cm-wave radio telescope -- SKA (see Gurvits this volume).

\subsection{Receiver Noise Temperature} 

Both cm and mm-VLBI are likely to see significant progress in the area
of improved receiver technology. For example, the $e$-MERLIN array (see
Muxlow et al. this volume) will employ 4-8~GHz receivers that are
expected to be a factor of two better than the current (33~K)
systems. Even larger factors of improvement can be expected at
mm-wavelengths (see Figure 2). 

 \begin{figure}
\vspace{8cm}  
\includegraphics{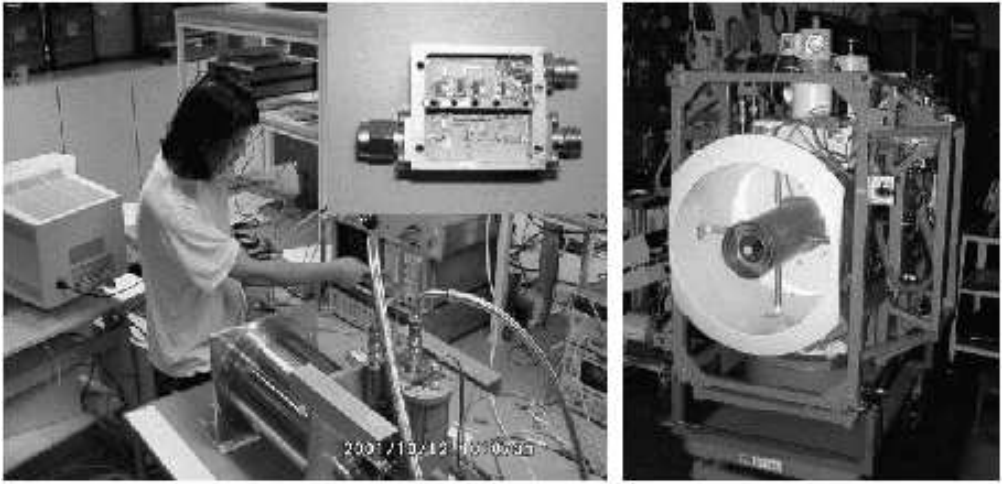}

\caption{Low-noise amplifiers (such as the broad-band 4-8 GHz 
  OSO/Chalmers system shown left) are set to make significant
  improvements in receivers to be developed for cm but in particular mm
  and sub-mm wavelengths. Low-noise frequency flexible systems (e.g.
  the WSRT MFFE system shown right) are now common at cm-wavelengths.}
\label{area} 
\end{figure}

\subsection{Coherence Time} 

\subsubsection{cm-VLBI}
Through the application of phase-referencing techniques (Beasley \&
Conway 1995), it is now possible to extend the coherence time of a
cm-VLBI array across the entire duration of the observations. This
permits the routine detection of relatively faint (sub-mJy) radio
sources (e.g. Garrett et al. 2001) at wavelengths up to $\lambda 1$~cm.
These phase-reference images are often dynamic range limited at the
level of 20:1, in-beam phase-referencing removes such limitations and
its use (at cm wavelengths) is generally more applicable than usually
understood. At $\lambda 18$~cm for example, in-beam phase-referencing
is even possible for very large telescopes, provided wide-field
techniques are used to combine the total signal from all compact
sources in the primary-beam (see section 7 for more details).
This permits {\it in-beam self-calibration} to be used in {\it all}
circumstances ({\it i.e.}  any target field) -- an important new
capability, and one that should be familiar to users of connected
arrays.

\subsubsection{mm-VLBI}
In recent years significant advances have been made in improving the
coherence times at mm-wavelengths. VERA for example will use a
multi-beam system that simultaneously permits one beam to be directed
at the target, and another to a nearby calibrator. Phase corrections
from the calibrator are thus continuously applied to the target without
any source switching. 

However, one limitation of dual-beam phase correction at mm wavelengths
is that the number of bright mm-calibrators is extremely limited, and
thus often involves target-calibrator separations on the scale of
several degrees. In these cases, the effects of spatial interpolation
errors (across the sky) cannot be ignored. Another method which addresses
this problem (and one which is showing enormous promise), is to derive
mm-wave phase corrections via a cm-wave reference source (see Asaki et
al.  1998; also Sasao et al., and Alef et al. this volume). This method
relies on the fact that the tropospheric delay is independent of
frequency. In its simplest form, multi-frequency co-axial feed systems
can be used to simultaneously observe the target, and phase
corrections derived at say 22~GHz are applied (after appropriate
scaling with frequency and correcting for any antenna based
phase-offsets) to much higher frequencies (e.g. 100 GHz). Note that
some relative astrometry is also preserved - the position of the
cm-wavelength ``core'' being the reference point.  It is clear that
this technique might also be usefully employed for high-frequency Space
VLBI missions, such as VSOP2 (see Haribyashi et al. this volume).

\subsection{Observing Bandwidth} 

The promise of access to much broader bandwidths is expected to explode
in the next decade. The consequences for both mm and cm-VLBI are
significant, especially for cm-VLBI where gains in other areas related
to sensitivity will be modest. Currently 256 Mbps is just about the
maximum data rate that can be currently sustained in most VLBI
networks. For the European VLBI Network (EVN, see {\tt www.evlbi.org}) the
limitation is not a technological one (512 Mbps recording is now
routinely performed) but the availability of thin-tapes. Within the
next 2 years sustained data rates of 1 Gbps will certainly be available
to VLBI users. This will lead to a factor of two better sensitivity for
both mm amd cm-VLBI networks. The longer term aim must be to attain
data rates of several Gbps, reaching tens of Gbps by the end of the
decade. This can be utilised by both mm and cm-VLBI networks, in the
latter case not just for bandwidth but in order to employ multi-bit
signal representation required by RFI mitigation algorithms.

\subsubsection{PC disk-based recorders} 

The promise of access to much broader bandwidths is expected to explode
in the next decade (see Figure 3). The consequences for both mm and
cm-VLBI are significant, especially for cm-VLBI where gains in other
areas will be modest.

The maximum total bandwidth currently used in VLBI is $\sim 64$~MHz (in
each of 2 polarisations), corresponding to a total data rate of 512
Mbps (2-bit signal representation and Nyquist Sampling). The
replacement of the current generation of magnetic tape recorders, with
PC-based disk recorder systems (see Whitney et al., Parsley et al.,
Romney et al., Kondo et al. this volume) is expected to take place over
the next few years. By employing commercial PC hardware, the VLBI
community will be able to take advantage of the rapid technological
development in this area. In principle, a doubling of the data capacity of
PC-based recorders might be expected every few years. Since the Mk5A
system can already record at 1 Gbps (see Figure 3), data rates in
excess of this are likely to be possible on relatively short time-scales.
In addition, the cost of disks will continue to shrink, at least in
real terms (i.e. for a given storage capacity).  

\subsubsection{Real Time, Optical Fibre-based VLBI networks} 

An alternative (or perhaps successor) to disk-based recorder systems is
the connection of VLBI networks via optical fibres (see Parsley \&
Whitney this volume \& Figure 3). Fibre communication networks are
ideally suited to the real-time transfer of huge amounts of data over
long distances. The adoption of direct fibre connections by $e$-MERLIN
(see Muxlow this volume) signals the progress that is being made in
this area. As the costs of these networks continue to fall, and as
commercial networks become more flexible, the introduction of a
real-time VLBI system is a reasonable goal to pursue. 

In Europe there are plans to demonstrate the feasibility of real-time
VLBI using shared IP routed networks (e.g.  G\'{E}ANT). A
proof-of-concept test programme (to be conducted over the next 1-2
years), aims to connect together directly, at least 4 European
telescopes to the EVN correlator at JIVE (see Parsley this volume).
Each telescope will generate up to 1 Gbps data streams, and these will
feed into the EVN correlator at JIVE (see Parsley et al. this volume).
The idea is to correlate the data with minimal buffering at either end
of the fibres. The provision of local loops (last mile connections) to
the telescopes and correlator is the critical item.  Local loops are in
place at Dwingeloo (JIVE) and Torun. Westerbork is to be connected in
mid-2003, and negotiations are on-going at other EVN sites, in
particular Jodrell Bank, Effelsberg, Medicina, Onsala and Metsahovi.

Technical issues that need to be thrashed out include the quality of
service required by VLBI, and the amount of buffering required at the
telescopes and correlator. Assuming the first tests are successful, the
ambition in Europe is to investigate ``production'' real-time VLBI
networks, and to broaden participation to include Asia, North-America
and Africa.

\begin{figure}
\vspace{12cm}  
\includegraphics{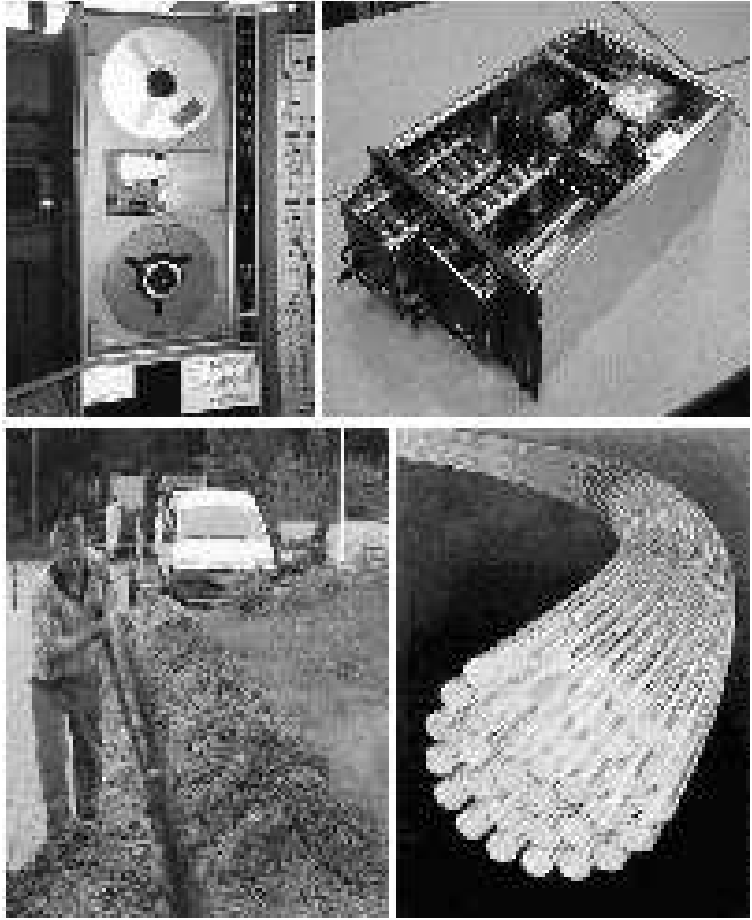}

\caption{Magnetic tape technology (top left) is now being replaced by PC
  disk-based (Mk5) recording systems (top right). Real-time VLBI using
  optical fibre networks must be the long-term goal (bottom right). The
  EVN correlator at JIVE is already connected by a fibre network that
  currently provide Gbps data rates but will be easily upgraded to
  permit even larger data rates to be employed.  }

\label{area} 
\end{figure}

\subsubsection{Implications for VLBI Back/Front-end systems \& Correlators} 

Both fibre and PC-based VLBI networks will permit routine Gbps VLBI
observations to be made in the course of the next few years.
Developments beyond this requires (in many cases) a replacement to the
current VLBI data acquisition system (in particular the expensive, and
now obsolete, analogue Base Band Converters - BBCs) with cheaper and
more flexible digital replacement systems. The interest in the latter
topic is witnessed by the activity reported in this volume (see Ferris
et al., Tuccari et al., Ying et al., Roh et al., Kondo et al., Koyama
et al., Iguchi et al). However, discussion about the necessity to
broad-band front-end telescope receiver systems was limited. 

Correlation of Gbps data stream is also a problem. For example, the EVN
MkIV correlator at JIVE can currently handle 16 telescopes at 1 Gbps or
(potentially) 8 telescopes at 2 Gbps. Data rates in excess of this
would require a new, more capable correlator, similar to that being
developed for the EVLA and $e$-MERLIN (see Carlson et al. this volume).

It is clear that to take full advantage of the increasing capacity of
both disk and fibre-based systems, considerable efforts must begin {\it
now}, in terms of new receiver systems, replacement back-end data
acquisition racks and future VLBI correlator developments.

\subsection{Overall Sensitivity Gains \& Image Fidelity} 

It is clear that both cm-VLBI and in particular mm-VLBI can expect to
make considerable gains in terms of collecting area, bandwidth (data
rates), receiver noise temperature and techniques to extend the
coherence time of the data. 

For continuum cm-VLBI a total gain in sensitivity of at least 5 seems
plausible over the next few years. In principle, noise levels at
microJy and even sub-microJy levels should be attainable (see Figure 4)
by Global VLBI arrays. An important provision is that both the EVN and
Very Long Baseline Array (VLBA, see {\tt www.nrao.edu}) adopt the same
fully compatible, next generation (disk-based) data acquisition
systems. It is good to see that the Global VLBI Working Group (that
also met here in Korea) have already started to worry about these very
issues.

In the case of mm-VLBI, at least an order of magnitude improvement
would not be surprising. In addition, since mm-VLBI's gains will also
include additional collecting area and improved receiver systems,
spectral-line studies also stand to benefit, not just standard
continuum observations.

\begin{figure}
\vspace{8cm}  
\includegraphics{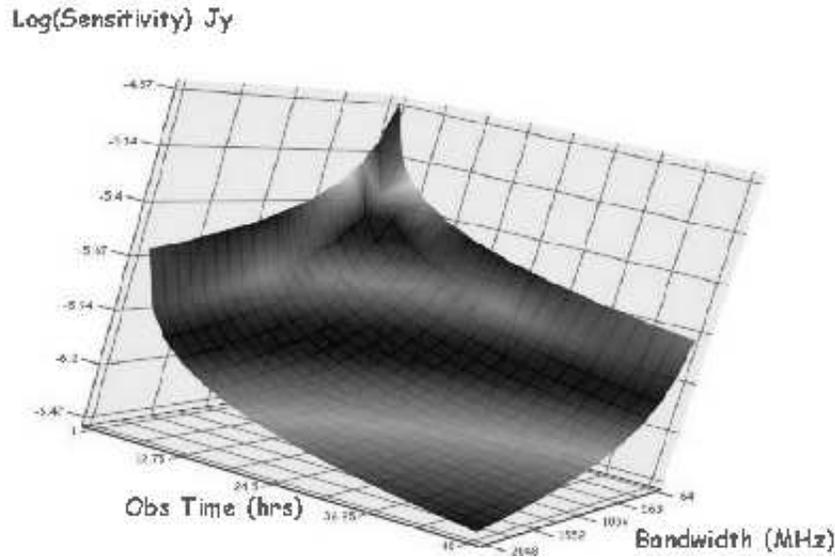}
\caption{The sub-microJy sensitivity of the $e$-EVN as a function of bandwidth and
  integration time.}
\label{hdf_zoom} 
\end{figure}

So far we have neglected to mention that any increase in observed
bandwidth will also (assuming the data remains unaveraged in frequency)
result in an improvement in uv-coverage and thus image fidelity. This
is particularly the case for cm-VLBI where the fractional bandwidth
should soon approach unity and Multi Frequency Synthesis (MFS)
techniques can be employed to take full advantage of this. Figure 5
presents the uv-coverage of the $e$-EVN assuming a total bandwidth of
2~GHz per polarisation.

\begin{figure}[h]
\vspace{6.0cm}  
\includegraphics{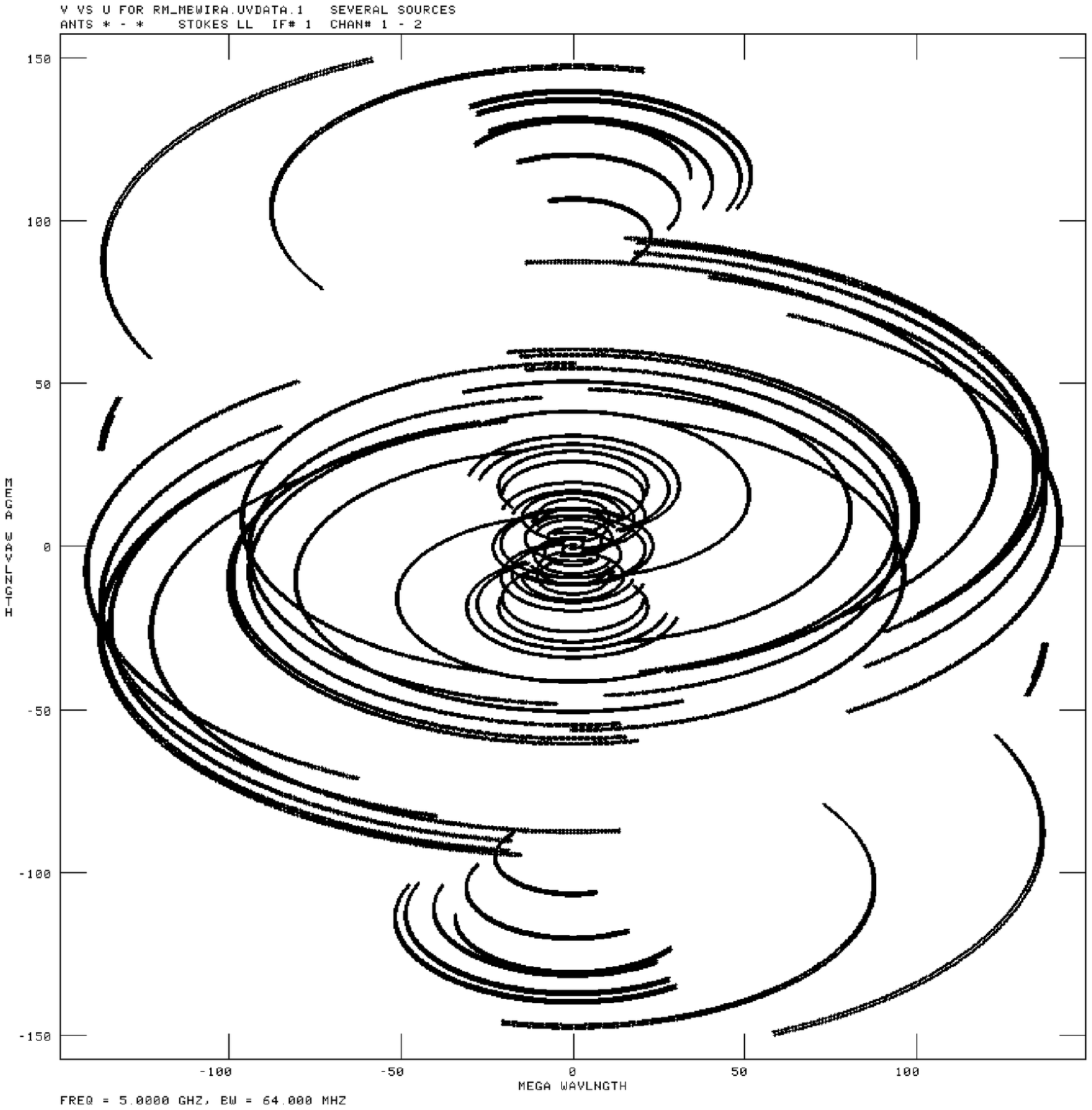}
\includegraphics{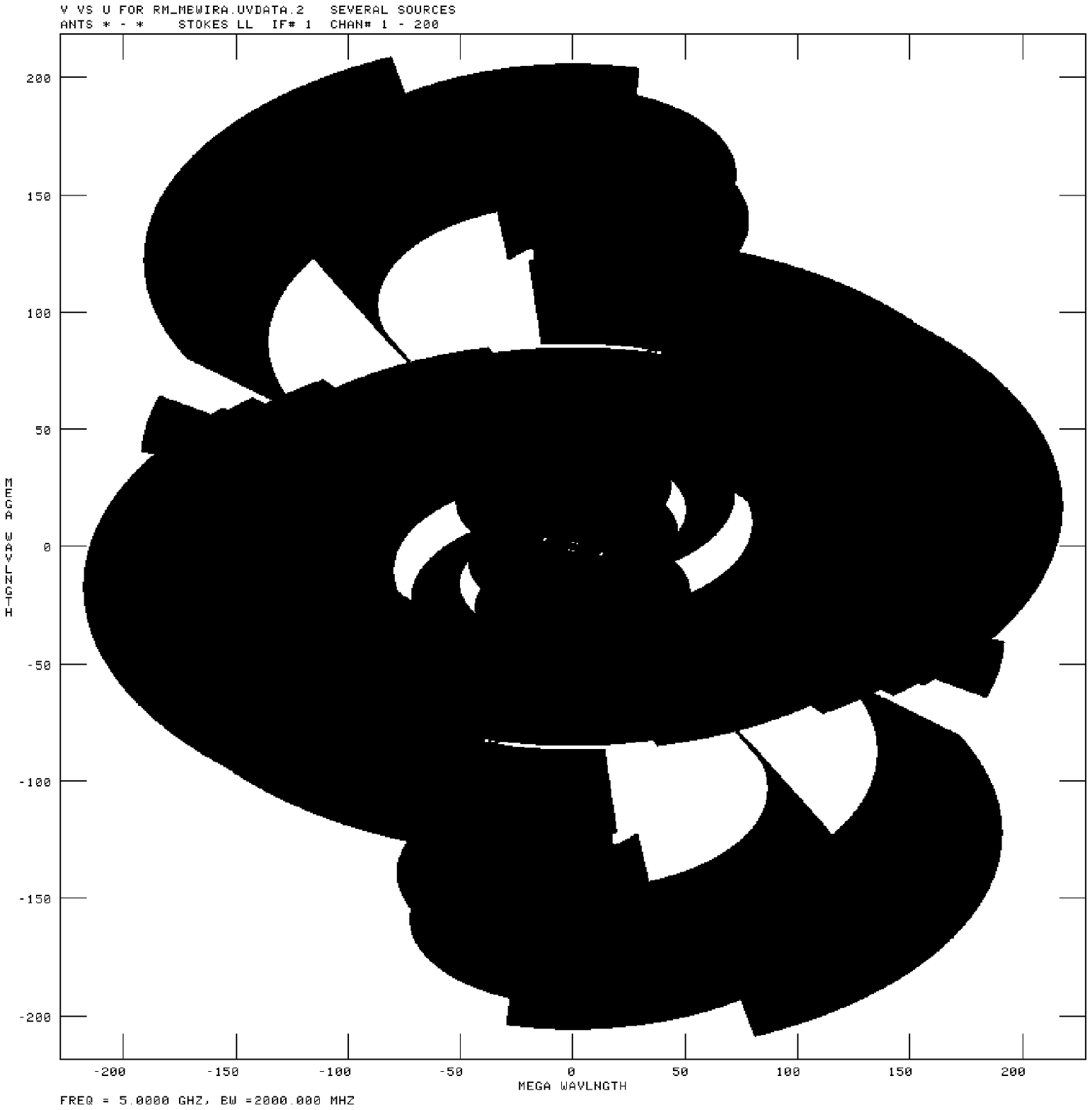}
\caption{Left: The current uv-coverage of the EVN at $\lambda6$~cm for a source
  located at $\delta=30^{\circ}$. Right: the extended (almost full)
  uv-coverage of the $e$EVN for the same source, assuming a total
  bandwidth of 2~GHz per polarisation.}
\label{uvcov} 
\end{figure}

\section{Frequency Coverage/Flexibility, Receiver Design \& RFI} 

An important development in recent times, has been the construction of
front-end receivers that can be instantaneously tuned over a wide-range
of sky frequency (e.g. from 1-10 GHz in the case of the Allen Telescope
Array). In these systems, several bands (each $\sim 1$~GHz in extent)
can then be selected individually and digitised. Systems like these
need to be in place at VLBI telescopes if we are to fully capitalise on
the expected increase in capacity of second generation disk and (first
generation) fibre-based VLBI networks.  However, in addition to
sensitivity and image fidelity issues (see previous section), there is
another astronomical motivation for instantaneous access to large
swathes of bandwidth: Serendipitous VLBI spectral line surveys (e.g. HI
in absorption).  Already such surveys are being conducted by connected
element arrays (e.g. Morganti \& Garrett 2002) and as the field-of-view
of VLBI observations increases (see section 6), VLBI can easily
follow suit.

The effects of Radio Frequency Interference (RFI) will also become
increasingly important as VLBI systems become more sensitive and
observe larger bandwidths. Although RFI does not usually correlate on
baselines $> 10$~km, local interfering signals are often so strong that
they can easily saturate receivers, and dominate the antenna system
noise. In addition, as a noise source, RFI is often extremely variable
on time scales of a few seconds or less --- tracking the telescopes
calibration under such conditions is usually impossible. Real VLBI
users are often sceptical of many RFI suppression techniques but at
this meeting there were several good presentations that suggest there
are more effective ways to counter RFI (e.g. Kesteven and Roshi this
volume) than simply deleting the data. We had better start using these
sooner, rather than later.

Rapid frequency switching is a routine observing mode for the VLBA.
This capability is important for spectral index mapping studies but
also increases the robustness and reliability of VLBI operations. Many
of the telescopes in the EVN are now frequency flexible (see Figure 2)
but only a few experiments have taken advantage of this facility so
far.  Rapid progress is expected to take place in this area over the
next year.

Finally, polarisation purity is another topic that often gets ignored
in VLBI receiver design. This is another area in which a homogeneous
array such as the VLBA has a significant advantage. The current aim of
the EVN is to produce $< 2$\% cross-talk between left and right hand
circular polarisation channels.  Although many of the EVN telescopes
are actually much better than this, some telescopes show cross-talk at
the level of $\sim 15$\%! These include special cases such as the WSRT
phased-array. These figures can limit the dynamic range of total
intensity images of even moderately bright radio sources, and for
polarisation studies, the level of impurity is large enough that second
order calibration corrections (usually assumed to be negligible) must
be accounted for.

\section{Sharper Resolution, Space VLBI and SKA Configurations} 

When it comes to resolution VLBI astronomers are a difficult lot to satisfy.
Their desire for increasingly better resolution forces them to move
towards higher frequencies and/or longer baselines. This is an effect
that was clearly demonstrated at this meeting. In particular, next
generation orbiting VLBI telescopes (e.g. VSOP-2) combine both these
elements together, employing only high-frequency receivers (8, 22 \&
43~GHz), and baselines between $3 - 5$ Earth radii (see Hirabayashi et
al. and Mochizuki et al.  Gurvits et al. this volume). RadioAstron
goes a step further with even longer baselines being proposed.

I feel it necessary to introduce a note of caution at this point. As we
move towards higher frequencies and longer baselines, AGN science
clearly stands to benefit. However, other areas of growing importance
are being neglected e.g. the study of SNe, SNR, micro-quasars, active
stars, HI absorption, OH emission, starburst galaxies, high-z
star-forming/AGN systems etc. These require high resolution too -
perhaps third generation Space VLBI missions will be able to address
these requirements too.

Angular resolution is also a hot topic in the discussion of array
configurations for the SKA (see Gurvits this volume). Plans for the
next generation of ground and space based astronomical observatories,
will provide much higher resolution (approaching traditional
milliarcsecond scales) for sub-mm, IR, optical/UV and x-ray
astronomers. This, together with source confusion is likely to see the
SKA deployed with complimentary baselines in excess of 1000~km (Garrett
et al. 2002).

\section{VLBI Calibration, auxiliary data products and automatic data
  (pipeline) analysis} 

Over the last 10 years the VLBA has the set the standard in terms of
generating accurate and homogeneous astronomical VLBI calibration data.
Meanwhile the rest of us have been playing ``catch-up''! The generation
of such data is vital in order to make VLBI transparent to all
astronomers (not just a few ``black-belt'' practitioners). In addition,
it is necessary in order to obtain high-dynamic range images via
both manual and automatic analysis paths (but especially the latter). 
 
The EVN is beginning to get there - continuous system temperature data
is now available, both as a function of time and frequency (see Figure
6). The calibration of the telescopes (via the NASA/GSFC Field System)
is considerably improved, with lots of essential new features (see
Himwich this volume). In addition, the EVN is now able to compare the
pointing position of the telescope and the direction of the target
source i.e.  it is now possible to generate ``flag files'' that can be
used to identify non-valid telescope data (see also Figure 6). Progress
in this area was also reported by other arrays (e.g. the LBA, Tingay
this volume).

\begin{figure}[h]
\vspace{4.6cm}  
\includegraphics{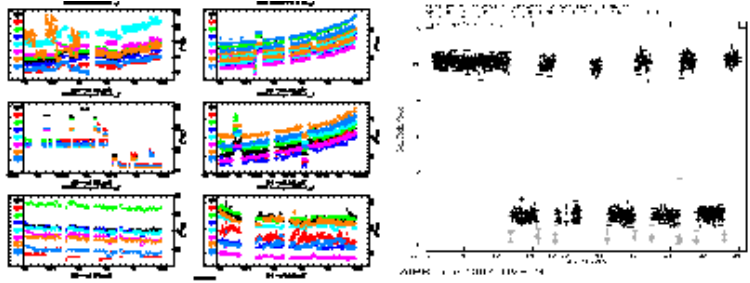}
\caption{In the 1990's the VLBA set the standard for auxiliary data
  products. Other VLBI networks are only now catching up. Continuous
  TSYS (left) and the application of Telescope Flag files (right) are
  now routinely generated by the EVN.}
\label{evn_im}
\end{figure}

As data sets become larger (see section 6 \& 7) it will be necessary
to automatically perform on-line calibration and data analysis. The EVN
pipeline (Reynolds, Paragi \& Garrett 2002) is the first step along
this road. All EVN projects are now ``pipelined'' by default. The
products include a set of AIPS calibration tables (a-priori calibration,
fringe-fitting and self-calibration) and various standard plots. As
well as reducing the effort required on behalf of the astronomer, we also gain
a much better understanding of the performance of the network. The
default mode is only to pipeline data associated with calibrators but
on request the target source can also be analysed. All astronomers can
take advantage of the EVN pipeline - irrespecitve of their experience, 
affiliation or geographical location.  

\section{Field-of-View, Spectral-line \& Fantastic Correlator output
  data rates} 

For a connected element array, the field-of-view is often set by the
primary beam size of the individual telescope elements. For VLBI this
is hardly ever the case. In VLBI, a more demanding limitation is set by
the fine spectral resolution and short integration times that must be
employed in order to circumvent {\it both} bandwidth smearing and time
averaging effects. Since preserving the field-of-view scales
(computationally) with baseline length {\it squared}, wide-field VLBI
analysis places enormous pressure on offline computer resources
(processing speed {\it and} disk space). These are many orders of
magnitude greater than for short-baseline, connected arrays. 

In the same vein, a VLBI field-of-view that is comparable with the
primary beam of the individual telescopes, places demands on the
correlator output data rate that are nothing short of ``fantastic''.
Pushed to their limits, current VLBI correlators are just about capable
of providing sufficient resolution in both time and frequency (e.g. 0.5
sec integration time and $1024\times62.5$~kHz channels) to permit the
inner 3 arcmin the telescope primary beam to be imaged out with full
sensitivity at 1.4 GHz. This corresponds to a correlator output data
rate ($\sim 1$~Mbyte/sec) - well short of what is required to map-out
the full (half-power point) primary beam of a 100-m, never-mind the
much larger field-of-view associated with 25 or 32-m class telescopes.

\begin{figure}[h]
\vspace{5cm}  
\includegraphics{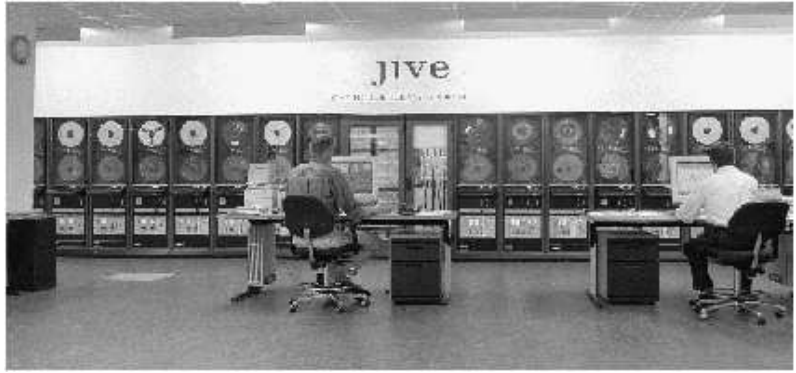}
\caption{The EVN correlator at JIVE has been operational since
  the end of 1999. The PCInt project will enable the EVN correlator to
  generate and handle output data rates as high as 160 MBytes/sec or 13
  TBytes/day. This will permit the full capacity of the correlator to
  be harnessed.  }
\end{figure}

The PCInt project currently being developed at JIVE (see Figure 7),
will enable the EVN correlator to generate and handle output data rates
as high as 160 MBytes/sec or 13 TBytes/day. This will permit the full
capacity of the correlator to be harnessed, permitting a spectral
resolution of 8092 channels per baseline or integration times as short
as 15 milli-seconds. PCInt will hugely expand the field-of-view of VLBI
quite generally (not just the EVN), and will provide the capability to
simultaneously map-out large swathes of the radio sky with
milliarcsecond resolution.

Much finer spectral resolution is not only required to expand the
field-of-view but it is also important for spectral-line studies.
Often spectral-line projects have to trade spectral resolution for the
number of telescopes, polarisation products etc. Many projects also
require multiple-pass correlation because more than one spectral
feature is present (and the correlator provides just enough resolution
for one line in any given pass). In addition, broad-banding of connected
interferometers (in particular the upgraded WSRT) has recently
revealed some very broad HI absorption systems (e.g. Morganti et al.
2002), broad enough, that current VLBI correlators are inadequate to
appropriately sample the full width of the line.

\section{Deep, Wide-Field cm-VLBI Studies} 

The application of wide-field techniques to VLBI data analysis is
fundamental to high resolution, deep field studies of the faint sub-mJy
radio source population. That VLBI can make a contribution in this area
was first demonstrated with the EVN 1.6~GHz observations of the Hubble
Deep Field (HDF). 

\subsection{EVN Observations of the HDF} 

EVN 1.6~GHz observations of the HDF (Garrett et al. 2001) were the
first VLBI observations of what is essentially a ``blank field'', i.e.
a region of sky, devoid of bright radio sources. The brightest source in
the HDF-N (as measured by the WSRT and VLA) is a 1.6~mJy FR1 radio
galaxy at $z=1$.  In addition, the observations (correlated at the VLBA
correlator in Socorro) employed wide-field techniques (1 sec
integrations, $64 \times 125$~kHz channels) and it was thus possible to
simultaneously image out the full field encompassed by the HDF-N ($\sim
6$~sq. arcmin). Three sources were detected within this field (see
Figure 8) - including the faintest source yet detected by VLBI - a 180
microJy AGN associated with a spiral galaxy (with a bulge) at $z=0.96$. 

\begin{figure}[h]
\vspace{13cm}  
\includegraphics{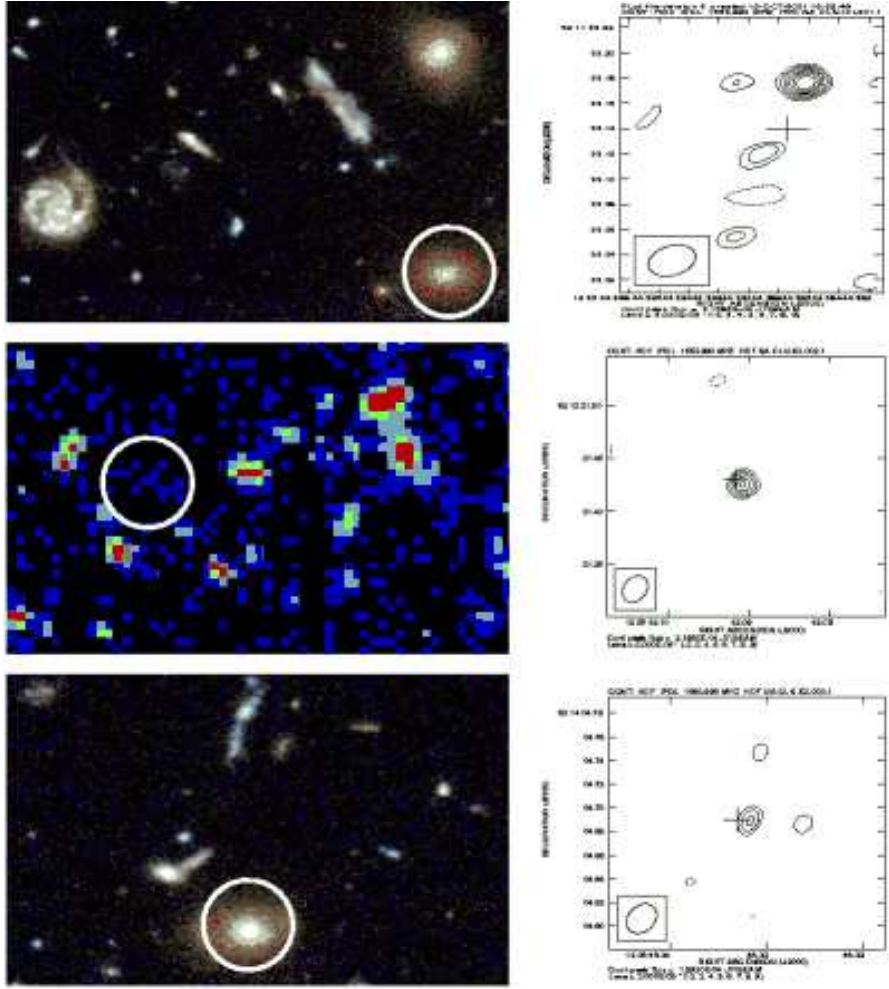}
\caption{EVN detections in the HDF: the distant
  z=1.01 FRI (top), the z=4.4 dusty obscured starburst hosting a
  hidden AGN (middle) and the faint 180 microJy,
  z=0.96 AGN (bottom). Crosses represent the MERLIN-VLA positions for
  these sources.}
\label{evn_im}
\end{figure}

\subsection{Recent VLBA+GBT deep field results} 

The rms noise levels achieved by the EVN HDF-N observations were
limited by phase errors introduced via conventional, external
phase-referencing (switching) techniques. The field-of-view was limited
by the frequency and time resolution that could then be achieved by the
VLBA correlator.

Some recent VLBA+GBT deep field observations illustrate the gains to be
made in employing ``in-beam'' phase referencing.  Figure 9 shows the
deepest VLBI images made to date (Garrett, Wrobel \& Morganti in prep).
The images (with an rms noise of 9 microJy/beam in the centre of the
field) were made from a 1.4~GHz VLBA+GBT observing run ($3\times8$
hours $@$256 Mbps). In-beam phase-referencing was used to provide
essentially perfect phase corrections for this data set, and eight
sources are simultaneously detected ($> 7\sigma$) within and outside
the half-power point of the GBT primary beam. Of these eight sources,
two sub-mJy sources are detected within the primary beam of the GBT, in
addition to the in-beam phase reference calibrator (a compact 20 mJy
source, first detected by Wrobel et al.). The images of sources far
from the field centred are tapered, since the time/frequency sampling
is only adequate for sources that lie within the primary beam of the
GBT (the latter being centred on the VLBI phase centre).

The total (target) data set size is 60 Gbytes (0.5 secs integration,
$1024 \times 62.5$~kHz channels).  Images were made with the AIPS task
IMAGR - dirty maps/beams of each sub-band (IF) for each epoch were
generated blindly and then simply co-added together.  Each postage
stamp image took about 8 hours to produce on a dual (2 GHz) processor
Linux box. The analysis of these data is on-going. For sources that
were bright enough, CLEAN maps were produced by simply subtracting the
dirty beam from the dirty image (AIPS task APCLN). More complicated
tasks (e.g. IMAGR) involving a visibility based CLEAN are prohibitively
expensive in terms of CPU requirements.

\begin{figure}
\vspace{12cm}  
\includegraphics{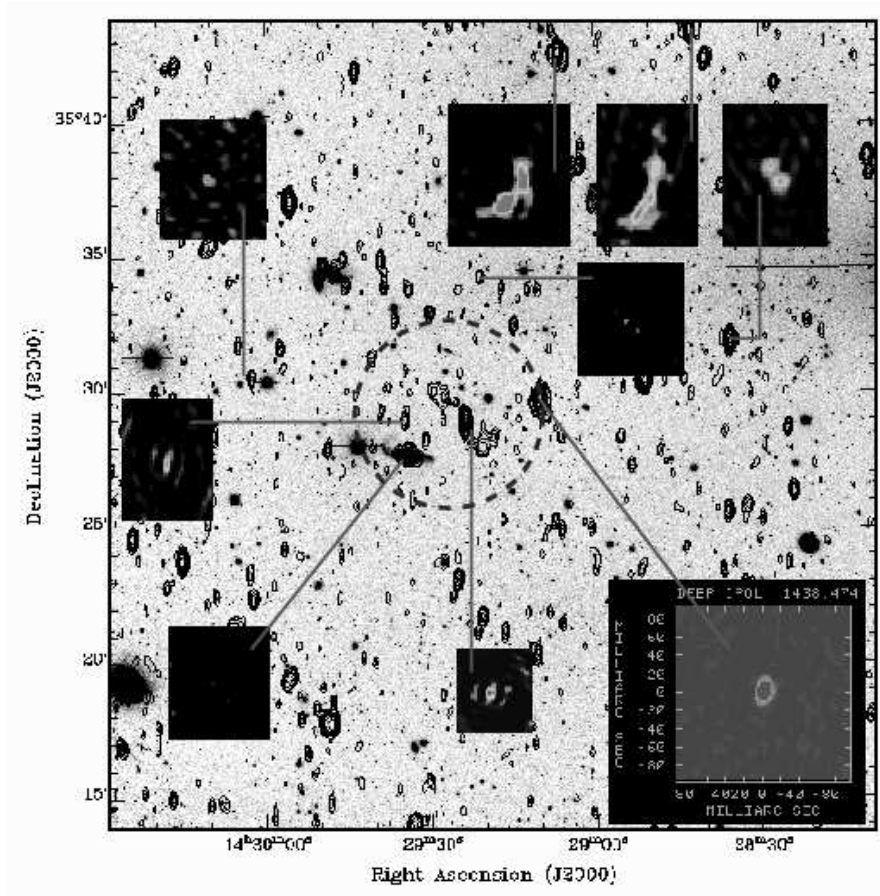}
\caption{Deep VLBA+GBT 1.4 GHz observations of a small portion of the
  NOAO-N Bootes deep field. The VLBI detections are shown inset. Radio
  line contours (produced by the WSRT) are superimposed on the NOAO
  optical field). One non-detection is also shown (bottom left) - a bright,
  presumably nearby spiral galaxy that is well detected by the WSRT
  (around the few mJy level). Very likely the radio emission from this system
  is associated with extended star formation. These are the deepest
  images made with VLBI to date (Garrett, Wrobel \& Morganti in
  preparation).  }
\end{figure}

\subsection{Lessons for the future}  

\subsubsection{Phase-referencing no longer required at 1-3~GHz}

What is clear now, and what I would like to focus on in the
final section of this paper, is that if the field of view of VLBI can
indeed be expanded (see section 6 and the PCInt development at
JIVE) then the whole concept of phase-referencing has to change - at
least at frequencies below 3~GHz.

The reason for this is that if one can simultaneously sample the {\it
  summed} response of all compact radio sources within (and indeed
beyond) the half-power point of the VLBI telescope primary beam, then
simple self-calibration of the target field is {\it always} possible,
essentially trivial, and can provide (essentially) perfect phase
corrections to the data.  Phase-referencing is still required (to some
extent) in order to preserve the astrometry and to improve the
coherence time of the data - {\it before} self-calibration of the
target field is attempted. In simple terms, at frequencies 1-3~GHz
(perhaps even higher or lower) a VLBI observation (correlated and
analysed using wide-field techniques) is pretty well like any connected
element array data set! Phase stability and coherence times are
essentially infinite.

From a technical perspective the message is clear: (i) With sustained
data rates of 1~Gbps, Global VLBI can approach, in some cases surpass,
the rms noise levels attained by connected element arrays and (ii)
every VLBI target field can be self-calibrated at frequencies $\sim 1-3
$~GHz, provided a wide enough field of view can be maintained, and sufficient
computing resources are available to cope with the enormous data sets
implied. 

Once systems like PCInt become available, the bottleneck quickly moves towards
the problem of handling the enormous data sets that such systems can generate.
Clusters of Linux PCs are the only feasible solution today, together
with massive storage devices. Access to GRID like computing resources may be
the only feasible short-term solution.

\section{A personal view of the future} 

The sensitivity of both mm and cm-VLBI is set to improve dramatically.
In the next 5-10 years, Gbps data rates will be routine, coherence
times will be virtually unlimited and the first real-time eVLBI
production networks will begin to be realised.  As network performance
is continuously monitored, serious telescope failures will be rare, and
feedback immediate. Target of Opportunity observations (e.g. GRBs) will
be possible even for part-time arrays such as the EVN. Frequency bands
will be configurable and blind searches for spectral features in the
data will be possible.  Wide-field imaging will be the norm, and
several dozens of sources will be detected and imaged simultaneously,
in any given observing run. Huge VLBI source surveys will be conducted
without relying on the biased selection criteria employed in major
surveys today. The resulting wide-field data sets will be usefully
mined by Virtual Observatory facilities, providing on-the-fly images
and spectra of particular regions of sky. 

The whole process of doing VLBI will be irevokobaly changed, and the
scientific base of our observations expanded immesureably. We can look
forward to an era in which, for the first time, VLBI observations will
be made, correlated and automatically pipelined all within the same
day! 

\begin{quote} 
\verb"I'd like to thank Young Chol Minh and the rest of the LOC for"\\
\verb"their hospitality and congratuate them on organising and"\\
\verb"hosting an excellent meeting". \\
\end{quote}

\section{References} 

\begin{quote} 
\verb"Asaki et al. 1998, Radio Science, 33, 1297."\\
\verb"Beasley, A. J., & Conway, J. E. 1995, ``VLBI "\\  "\\
\verb"Phase-Referencing,'' in ASP Conf. Series 82, Very "\\
\verb"Long Baseline Interferometry and the VLBA, ed."\\ 
\verb"J. A. Zensus, P. J. Diamond, & P. J. Napier"\\
\verb"(San Francisco: ASP), 327-343."\\ 
\verb"Garrett, M.A., Muxlow, T.W.B., Garrington, S.T. et
al."\\ \verb"2001, A&A Letters, 366, L5 (astro-ph/0008509)."\\
\verb"Garrett, M.A. 2002, in Procs. of the 6th EVN Symposium,"\\
\verb"ed. Ros, E. et al. (astro-ph/0211013)."\\
\verb"Morganti, R. & Garrett, 2002, M.A. WSRT Newsletter No. 17, p6."\\
\verb"Morganti, R. et al. 2002, in Proceedings of the 3rd GPS/CSS"\\
\verb"Workshop, ed. T.  Tzioumis, W.de Vries, I. Snellen,"\\
\verb" & A. Koekemoer (astro-ph/0212321)." \\
\verb"Reynolds, R. Paragi, Z. \& Garrett, M.A. 2002, Presented at"\\
\verb"the URSI General Assembly (astro-ph/0205118)."\\
\verb"Walker, R.C., 1988, in VLBI Techniques & Applications"\\
\verb"ed. Felli & Spencer, Kluwer Academic Publishers. "\\
\end{quote}

\end{document}